\documentclass{iau} 
\usepackage{graphicx}
\usepackage{caption}
\def\HII{H\,\textsc{ii}}

\title[JD 11.~~Planetary Nebula Candidates Uncovered with the HASH Research Platform]
{Planetary Nebula Candidates Uncovered with the HASH Research Platform}

\author[Fragkou V. et al.] 
{Vasiliki Fragkou$^{1,2}$, Ivan Boji\v{c}i\'c$^{1,2}$, David Frew$^{1,2}$, Quentin Parker$^{1,2}$}

\affiliation{$^1$The University of Hong Kong, Department of Physics, Hong Kong SAR, China\\
$^2$The University of Hong Kong, Laboratory for Space Research, Hong Kong SAR, China
 \\ email: {\tt vfrag@hku.hk} 
}

\pubyear{2017}
\volume{323}  
\jname{Planetary Nebulae: Multi-Wavelength Probes of Stellar and Galactic Evolution}
\editors{Xiaowei Liu, Letizia Stanghellini, \&  Amanda Karakas, eds.}
\begin{document}

\maketitle

\begin{abstract}
A detailed examination of new high quality radio catalogues (e.g. Cornish) in combination with available mid-infrared (MIR) satellite imagery (e.g. Glimpse) has allowed us to find 70 new planetary nebula (PN) candidates based on existing knowledge of their typical colors and fluxes. To further examine the nature of these sources, multiple diagnostic tools have been applied to these candidates based on published data and on available imagery in the HASH (Hong Kong/ AAO/ Strasbourg H$\alpha$ planetary nebula) research platform. Some candidates have previously-missed optical counterparts allowing for spectroscopic follow-up. Indeed, the single object spectroscopically observed so far has turned out to be a bona fide PN.
\end{abstract}

\firstsection %
\section{Introduction}

Modern multi-wavelength imaging surveys can help identify planetary nebulae hidden at optical wavelengths due to extinction (\cite[Cohen et al. 2011]{Coh_etal11}). The Cornish catalogue, which is based on the 5\,GHz radio continuum Cornish survey (\cite[Hoare et al. 2012]{Hoa_etal12}; \cite[Purcell et al. 2013]{Pur_etal13}), covers the northern GLIPMSE region (\cite[Benjamin et al. 2003]{Ben_etal03}; \cite[Churchwell et al. 2009]{Chu_etal09}), providing us with a new tool for the detection of PN candidates.

\section{Selection Method and Diagnostic Tools}

\begin{figure}[h]
\vspace*{-0.24 cm}
\begin{center}
\includegraphics[width=3.4in]{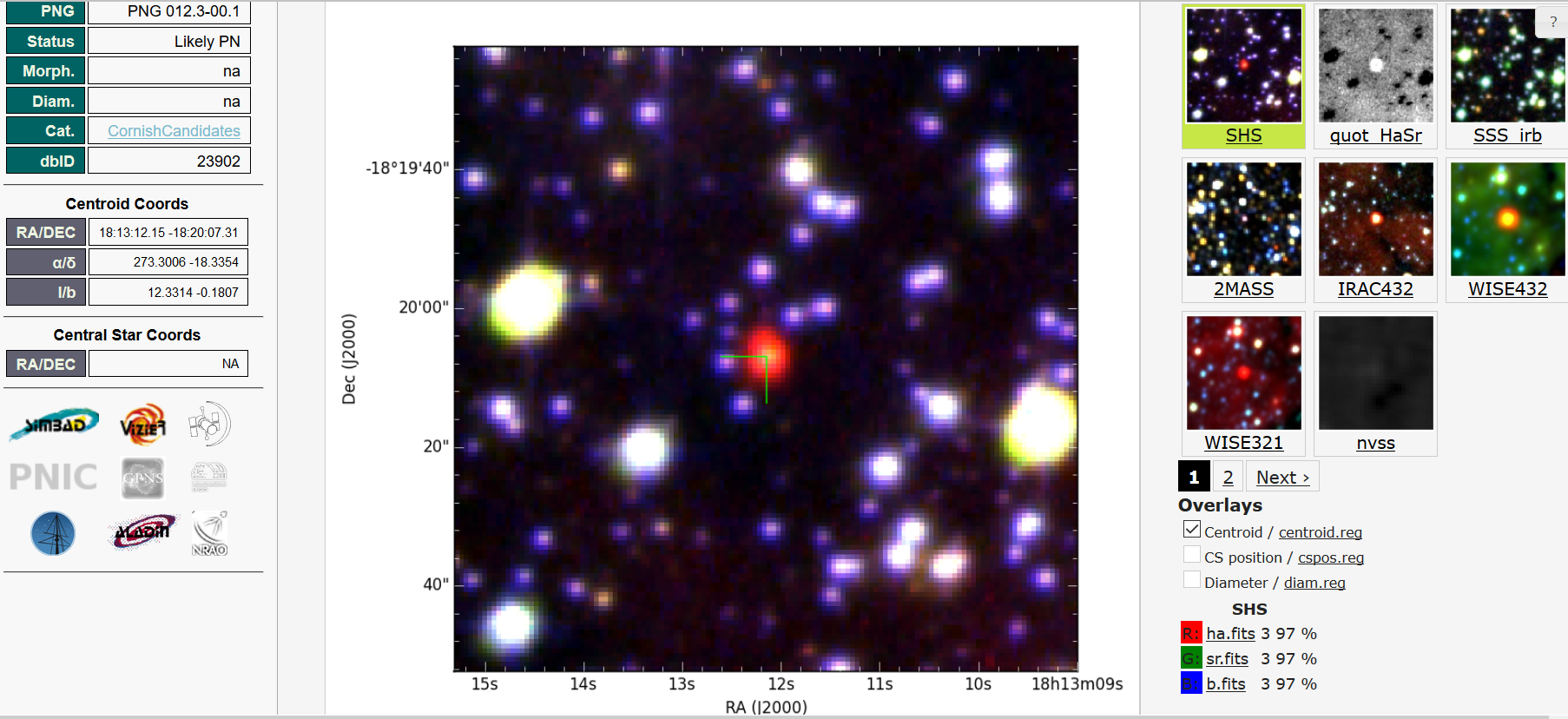} 
 \vspace*{-0.24 cm}
 \caption{Example of a new PN discovered by this study, shown in HASH  (Parker et al. 2016).
}
   \label{Fig 1}
\end{center}
\end{figure}

After cross-correlating the Cornish (\cite[Hoare et al. 2012]{Hoa_etal12}; \cite[Purcell et al. 2013]{Pur_etal13}), NVSS (\cite[Condon et al. 1998]{Con_etal98}) and Glimpse (\cite[Benjamin et al. 2003]{Ben_etal03}; \cite[Churchwell et al. 2009]{Chu_etal09}) catalogues, we rejected objects with $S_{\rm 5GHz}$  fluxes, \textgreater 110 mJy or a spectral index $\alpha$\,\textless$ -0.5$ following \cite[Anderson et al. (2011)]{And_etal11}. We also included in our sample objects with no NVSS counterpart. After visual inspection of multi-wavelength images using the HASH database (\cite[Parker et al. 2016]{Pabofr_16}, and these proceedings), we uncovered 70 PN candidates, 21 of which have an optical detection, including our first confirmed PN (see Fig.\,\ref{Fig 1}).

\begin{figure}[h]
\vspace*{-0.25 cm}
\begin{center}
\includegraphics[width=3.7in]{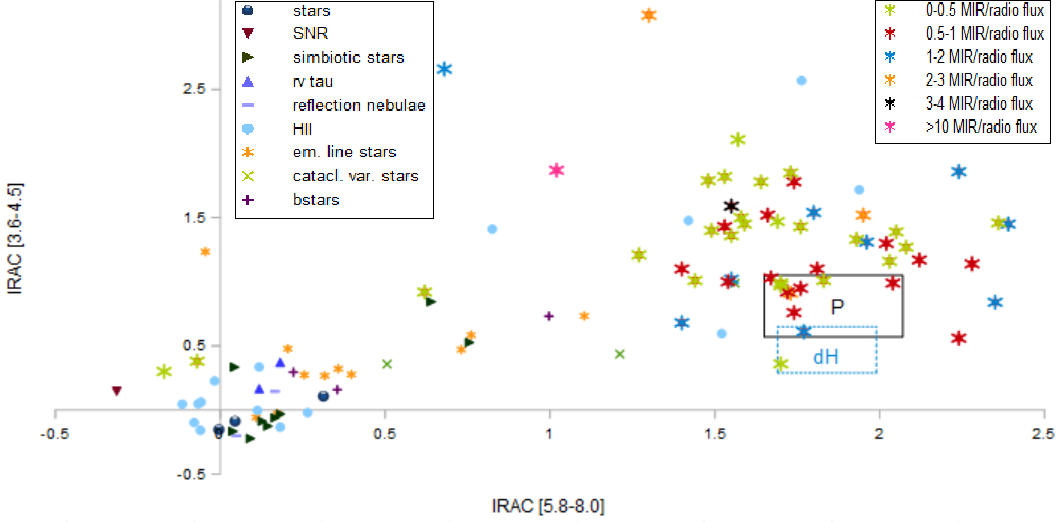} 
 \vspace*{-0.25 cm}
 \caption{IRAC color-color plot of different object types.
}
   \label{Fig 2}
\end{center}
\end{figure}

Diagnostic tools using both emission-line and continuum fluxes from multiple wavelengths can be used to separate PNe from \HII\ regions and non-thermal emitters (\cite[Cohen \& Green 2001]{Cogr_01}; \cite[Frew \& Parker 2010]{FP10}; \cite [Parker et al. 2012]{Par_etal12}; \cite[Frew et al. 2014]{Frew_etal14}). Previously known PNe have a median value of $F_{\rm 8\mu m}/S_{\rm 843MHz}= 4.7  \pm 1.1$, while \HII\  regions and non-thermal emitters present MIR/radio ratios of $\approx$25  and $\approx$0.06 respectively (\cite[Cohen et al. 2007]{Coh_etal07}; \cite[Cohen et al. 2011]{Coh_etal11}). Assuming that all of our PN candidates are optically thin, we can convert their $S_{\rm 5GHz}$ fluxes to $S_{\rm 843MHz}$ by using the equation $S_{\rm5GHz}/S_{\rm843MHz} =(5/0.843)^{-0.1}$ (\cite[Anderson et al. 2011]{And_etal11}).  In Figure\,\ref{Fig 2} we can see the MIR/radio flux ratios of our PNe candidates overplotted on an IRAC color-color plot. The black and blue boxes indicate the areas where most known PNe and \HII\  regions are placed, based on published data (\cite[Cohen et al. 2011]{Coh_etal11}). Data for other objects are obtained from \cite[Kurtz et al. (1994)]{Kur_etal94}, \cite[Giveon et al. (2005)]{Giv_etal05} and Parker et al. (2012, 2016).

\section{Results}

As shown in Fig.\,\ref{Fig 2}, most of our candidates are plotted on or around the box where most known PNe are located. Moreover, the MIR/radio flux ratios of most candidates agree well with previously known PNe, indicating there is a high probability of our candidates being true PNe. Follow-up studies are needed to confirm this hypothesis, especially for the 21 objects (30\% of the total) that have optical counterparts allowing spectroscopic follow-up. At the time of writing, one of these has already been confirmed as a bona fide PN (Fig. 1). Future work should also investigate the sources with $S_{\rm 5GHz}$ \textgreater 110 mJy.


\begin{thebibliography}{}

\bibitem[\protect\citeauthoryear{Anderson et al.}{2014}]{And_etal14} Anderson L.~D., et al., 2014, ApJS, 212, 1 

\bibitem[\protect\citeauthoryear{Anderson et al.}{2011}]{And_etal11} Anderson L.~D., Bania T.~M., Balser D.~S., Rood R.~T., 2011, ApJS, 194, 32 

\bibitem[\protect\citeauthoryear{Benjamin et al.}{2003}]{Ben_etal03} Benjamin R.~A., et al., 2003, PASP, 115, 953 

\bibitem[\protect\citeauthoryear{Churchwell et al.}{2009}]{Chu_etal09} Churchwell E., et al., 2009, PASP, 121, 213 

\bibitem[\protect\citeauthoryear{Cohen \& Green}{2001}]{Cogr_01} Cohen M., Green A.~J., 2001, MNRAS, 325, 531 

\bibitem[\protect\citeauthoryear{Cohen et al.}{2007}]{Coh_etal07} Cohen M., et al., 2007, MNRAS, 374, 979 

\bibitem[\protect\citeauthoryear{Cohen et al.}{2011}]{Coh_etal11} Cohen M., et al., 2011, MNRAS, 413, 514 

\bibitem[\protect\citeauthoryear{Condon et al.}{1998}]{Con_etal98} Condon J.~J., et al., 1998, AJ, 115, 1693 

\bibitem[\protect\citeauthoryear{Condon \& Parker}{2010}]{FP10} Frew D.~J., Parker Q.~A., 2010, PASA, 27, 129 

\bibitem[\protect\citeauthoryear{Frew et al.}{2014}]{Frew_etal14} Frew D.~J., et al., 2014, MNRAS, 440, 1345   

\bibitem[\protect\citeauthoryear{Giveon et al.}{2005}]{Giv_etal05} Giveon U., Becker R.~H., Helfand D.~J., White R.~L., 2005, AJ, 130, 156 

\bibitem[\protect\citeauthoryear{Hoare et al.}{2012}]{Hoa_etal12} Hoare M.~G., et al., 2012, PASP, 124, 939 

\bibitem[\protect\citeauthoryear{Kurtz, Churchwell, \& Wood}{1994}]{Kuchwo_94} Kurtz S., Churchwell E., Wood D.~O.~S., 1994, ApJS, 91, 659 

\bibitem[\protect\citeauthoryear{Parker et al.}{2012}]{Par_etal12} Parker Q.~A., et al., 2012, MNRAS, 427, 3016

\bibitem[\protect\citeauthoryear{Parker, Boji{\v c}i{\'c}, \& Frew}{2016}]{Pabofr_16} Parker Q.~A., Boji{\v c}i{\'c} I.~S., Frew D.~J., 2016, JPhCS, 728, 032008 

\bibitem[\protect\citeauthoryear{Purcell et al.}{2013}]{Pur_etal13} Purcell C.~R., et al., 2013, ApJS, 205, 1 

\end{thebibliography}
\end{document}